\documentclass[aps,prb,reprint,twocolumn,floatfix,superscriptaddress,amsmath,amssymb,amsfonts, nofootinbib]{revtex4-2}

\usepackage{url}
\usepackage{bm}
\usepackage{graphicx}
\usepackage{amsmath}
\usepackage{amstext}
\usepackage{amssymb}
\usepackage{amsfonts}
\usepackage{amsbsy}
\usepackage{verbatim}
\usepackage{color}
\usepackage{braket}
\usepackage[colorlinks=true, urlcolor=blue, linkcolor=blue, citecolor=blue, pdftex]{hyperref}
\usepackage{floatrow}
\usepackage{float}
\usepackage{tikz}
\usepackage{gensymb}
\usepackage{textcomp}
\usepackage{enumitem}
\usepackage[version=3]{mhchem}
\usepackage{afterpage}
\usepackage{pbox}
\usepackage{dsfont}
\usepackage{siunitx}
\usepackage{stackengine,wasysym,scalerel}
\usepackage{latexsym}
\usepackage{cancel}
\usepackage{comment}
\usepackage{array, multirow, bigdelim, makecell, booktabs}
\usepackage[misc]{ifsym}
\usepackage[capitalise]{cleveref}
\usepackage{times}

\usepackage{color}
\usepackage[colorlinks=true, urlcolor=blue, linkcolor=blue, citecolor=blue, pdftex]{hyperref}

\usepackage[normalem]{ulem}

\definecolor{darkblue}{HTML}{004D6B}
\definecolor{darkred}{HTML}{8c1515}
\definecolor{darkgreen}{HTML}{006400}
\usepackage{hyperref}
\hypersetup{
	colorlinks=true,
	urlcolor=darkblue,
	citecolor=darkblue,
	linkcolor=darkblue,
	breaklinks
}

\newcommand{\dagga}{{\phantom{\dagger}}}

\begin{document}

\title{Monopole excitations in the $U(1)$ Dirac spin liquid on the triangular lattice}

\author{Sasank Budaraju}
\affiliation{Laboratoire de Physique Th\'eorique, Universit\'e de Toulouse, CNRS, UPS, France}
\affiliation{Department of Physics and Quantum Centre of Excellence for Diamond and Emergent Materials (QuCenDiEM), Indian Institute of Technology Madras, Chennai 600036, India}

\author{Alberto Parola}
\affiliation{Dipartimento di Scienza e Alta Tecnologia, Universit\`a dell'Insubria, Via Valleggio 11, I-22100 Como, Italy}

\author{Yasir Iqbal}
\affiliation{Department of Physics and Quantum Centre of Excellence for Diamond and Emergent Materials (QuCenDiEM), Indian Institute of Technology Madras, Chennai 600036, India}

\author{Federico Becca}
\affiliation{Dipartimento di Fisica, Universit\`a di Trieste, Strada Costiera 11, I-34151 Trieste, Italy}

\author{Didier Poilblanc}
\affiliation{Laboratoire de Physique Th\'eorique, Universit\'e de Toulouse, CNRS, UPS, France}

\date{\today}

\begin{abstract}
The $U(1)$ Dirac spin liquid might realize an exotic phase of matter whose low-energy properties are described by quantum electrodynamics in $2+1$ dimensions, where 
gapless modes exists but spinons and gauge fields are strongly coupled. Its existence has been proposed in frustrated Heisenberg models in presence of frustrating 
super-exchange interactions, by the (Abrikosov) fermionic representation of the spin operators [X.-G. Wen, \href{https://doi.org/10.1103/PhysRevB.65.165113}{Phys. 
Rev. B {\bf 65}, 165113 (2002)}], supplemented by the Gutzwiller projection. Here, we construct charge-$Q$ monopole excitations in the Heisenberg model on the 
triangular lattice with nearest- ($J_1$) and next-neighbor ($J_2$) couplings. In the highly frustrated regime, singlet and triplet monopoles with $Q=1$ become 
gapless in the thermodynamic limit; in addition, the energies for generic $Q$ agree with field-theoretical predictions, obtained for a large number of gapless 
fermion modes. Finally, we consider localized gauge excitations, in which magnetic $\pi$-fluxes are concentrated in the triangular plaquettes (in analogy with 
$\mathbb{Z}_2$ visons), showing that these kind of states do not play a relevant role at low energies. All our findings lend support to a stable $U(1)$ Dirac spin 
liquid in the $J_1-J_2$ Heisenberg model on the triangular lattice.
\end{abstract}

\maketitle

\section{Introduction}

Quantum spin liquids are exotic states of matter that escape any spontaneous symmetry breaking, down to zero temperature, and feature emergent fractionalized degrees 
of freedom~\cite{savary2017}. As a consequence, they cannot be characterized by local order parameters, thus lying beyond the Landau paradigm, but instead possess 
non-trivial patterns of long-range entanglement and topological orders. An increasing evidence for their existence comes from recent theoretical, numerical, and 
experimental studies~\cite{broholm2020}. The physics of these compounds is primarily governed by Heisenberg-like interactions between the local spin moments. 
In general, the strategy to stabilize spin liquids comes from the presence of competing (i.e., frustrating) super-exchange couplings, as exemplified in the original 
proposal for the nearest-neighbor Heisenberg model on the triangular lattice~\cite{anderson1973,fazekas1974} (even though it is now well accepted that the ground 
state of this model is magnetically ordered~\cite{capriotti1999,white2007}). Nevertheless, the inclusion of a next-nearest-neighbor term (thus leading to the 
$J_1-J_2$ Heisenberg model) may lead to a genuine spin-liquid phase, which could be either gapped~\cite{zhu2015,hu2015} or gapless~\cite{kaneko2014,iqbal2016,hu2019}.

The detection and characterization of quantum spin liquids is also important in connection to the possible emergence of (high-temperature) superconductivity upon 
doping~\cite{anderson1987,baskaran1993,wen2006}, leading to enormous ongoing efforts to realize these states experimentally~\cite{norman2016}. In particular, for 
the triangular lattice, several candidate spin-liquid compounds have been proposed over the years, and elucidating their low-energy propeties is still an active 
area of research. For example, the spin-orbit-coupled insulator YbMgGaO$_4$ has garnered significant attention as a potential spin liquid~\cite{li2015}, even 
though alternative proposals for a spin glass phase have been proposed~\cite{ma2018}. Sodium chalcogenides NaYbX$_2$ (with X=S,O,Se)~\cite{bordelon2019} or
organic salts such as $\kappa$(ET)$_2$Cu$_2$(CN)$_3$~\cite{yamashita2008} and EtMe$_3$Sb[Pd(dmit)$_2$]$_2$~\cite{yamashita2008,yamashita2010} also showed some
evidences in this direction. More recently, the YbZn$_2$GaO$_5$ was argued to host a $U(1)$ Dirac quantum spin liquid~\cite{xu2023}.

An instructive approach to describe spin liquids is provided by parton constructions, where the spin operators are represented in terms of auxiliary fermionic 
or bosonic degrees of freedom~\cite{baskaran1988,arovas1988,affleck1988}. Here, the original Hilbert space is enlarged and gauge fields appear; the resulting 
Hamiltonian is strongly interacting, making any perturbative approach futile~\cite{zhou2017}. Nevertheless, it has been argued that a mean-field approximation, 
in which gauge fields are frozen in specific configurations, may give insightful information~\cite{wen2002,wenbook}. These are the cases in which the low-energy 
gauge fluctuations have $\mathbb{Z}_2$ symmetry, for which the mean-field properties are expected to be qualitatively preserved. Instead, the $U(1)$ cases are 
much more subtle~\cite{hermele2004}; here, the major complication comes from the existence of magnetic monopoles, which correspond to instanton events and may 
proliferate, thus invalidating the mean-field picture. In recent years, several studies have explored monopoles and their role in the stability of the Dirac spin 
liquid~\cite{dupuis2019,nambiar2023,seifert2024,ganesh2024,dumitrescu2024}. The condensation of triplet and singlet monopoles is expected to generate antiferromagnetic or 
valence-bond order, respectively~\cite{hermele2004}. However, in some frustrated lattices, a recent field-theoretical analysis has shown that monopoles have 
non-trivial quantum numbers (i.e., they are not in the total-symmetric sector) and do not lead to such instabilities in fermionic models with a sufficiently 
large number of gapless (Dirac) points~\cite{song2019,song2020}. These results strongly suggest that a gapless phase, where fermions and $U(1)$ gauge fields are 
strongly coupled (e.g., the quantum electrodynamics in $2+1$ dimensions), may be realized in microscopic Heisenberg models on frustrated lattices, as previously 
suggested by intensive numerical investigations~\cite{kaneko2014,iqbal2016,hu2019,iqbal2013,he2017}. In particular, a recent work on the $J_1-J_2$ model on the 
triangular lattice compared the exact low-energy spectrum on small clusters (e.g., up to $48$ sites) with the one obtained within a parton construction with 
$U(1)$ gauge fields and gapless fermions, suggesting the possibility that the magnetically disordered region realizes a gapless $U(1)$ spin liquid~\cite{wietek2024}. 
Pursuing this kind of analysis on larger clusters is fundamental to shed light into the properties of highly-frustrated magnets and their possible description in 
terms of fermionic and gauge degrees of freedom.

In this work, we focus on monopole excitations within Gutzwiller-projected wave functions and assess their energetics for the $J_1-J_2$ Heisenberg model in the 
highly-frustrated regime, where previous numerical calculations have pushed forward the existence of a gapless spin-liquid phase~\cite{kaneko2014,iqbal2016,he2017}. 
The construction of variational wave functions relies on the parton approach, supplemented by the Gutzwiller projector, which allow us to work in the correct Hilbert
space of the spin model. In the highly-frustrated regime ($J_2/J_1 \approx 1/8$), the mean-field (quadratic) Hamiltonian has hopping terms only, with a particular
choice of signs that leads to non-trivial magnetic fluxes~\cite{iqbal2016}. As a result, the fermionic spectrum possesses a couple of Dirac points in the Brillouin 
zone (for each spin value), thus giving $N_{f}=4$ gapless fermions in total. Then, charge-$Q$ monopoles are constructed by changing the fermionic hoppings, so as 
to insert a $2\pi Q$ flux, uniformly distributed in the whole lattice. For $Q=1$, we observe gapless excitations in the thermodynamic limit, similar to what has been 
detected on the kagome lattice~\cite{budaraju2023}. Remarkably, the overlap between monopole states (both singlet and triplet) and particle-hole spinon excitations 
(as considered in Ref.~\cite{ferrari2019}, obtained without modifying the hopping parameters) are large on the $6 \times 6$ cluster, suggesting that spinons and 
monopoles are strongly coupled. Monopoles with charge-$Q$ ($Q>1$) are also considered, and their energetics agree well with the field-theoretical calculations 
obtained in the limit of a large number of gapless fermions~\cite{borokhov2002,dyer2013,dupuis2022}. Finally, we discuss the role of localized gauge excitations, 
where a magnetic $\pi$ flux is fully concentrated in a single triangular plaquette, as for $\mathbb{Z}_2$ visons. In contrast to the latter case, where these states 
represent elementary excitations~\cite{tay2011}, here, they have a large overlap with the ground-state wave function and do not represent low-energy excitations.

The paper is organized as follows: in Sec.~\ref{sec:model}, we describe the $J_1-J_2$ Heisenberg model and the variational wave function for the $U(1)$ 
spin liquids; in Sec.~\ref{sec:symmetry}, we discuss important aspects about the symmetries of the ground-state and monopole wave functions; numerical results 
are reported in Sec.~\ref{sec:results}, and, finally, conclusions are drawn in Sec.~\ref{sec:conclusions}.

\begin{figure}
\includegraphics[width=\columnwidth]{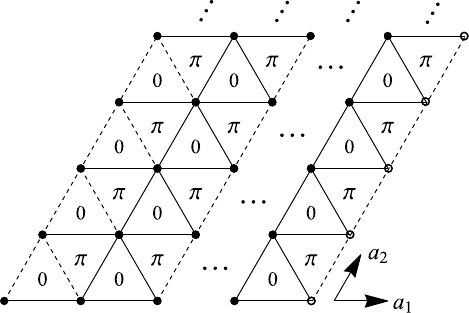}
\caption{\label{fig:LxLdirac}
Hopping amplitudes $\chi_{ij}$ for the Dirac spin liquid on a $L \times L$ cluster. Solid and dashed lines denote the hoppings $\chi_{ij}=1$ and $-1$, respectively.
The corresponding magnetic fluxes ($0$ and $\pi$) are also reported.}
\end{figure}

\begin{figure}
\includegraphics[width=\columnwidth]{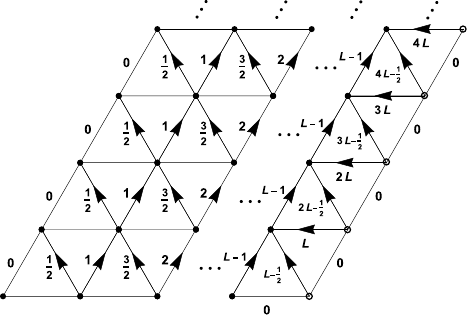}
\caption{\label{fig:LxLmonopole}
Pattern of the phases of hoppings $\theta_{ij}$ for the uniform monopole {\it Ansatz} on a $L\times L$ cluster. The numbers indicate the values of the phases 
in units of $2\pi/N$. Each plaquette has a flux of $2\pi/N$. The total hopping amplitude is given by $\chi_{ij} e^{i\theta_{ij}}$, where $\chi_{ij}=\pm 1$ is the 
underlying pattern of the Dirac state. The last column of hoppings result in translation symmetry being broken along both lattice directions.}
\end{figure}

\section{Model and parton construction}\label{sec:model}

In this work, we consider the Hamiltonian of interacting $S=1/2$ spins on the triangular lattice:
\begin{equation}\label{eq:ham}
{\cal H} = J_1 \sum_{\langle i j \rangle} {\bf S}_i \cdot {\bf S}_j + J_2 \sum_{\langle \langle i k \rangle \rangle} {\bf S}_i \cdot {\bf S}_k,
\end{equation}
where ${\bf S}_i=(S^x_i,S^y_i,S^z_i)$ is the spin 1/2 operator on site $i$ and $\langle ij\rangle$ ($\langle \langle ik \rangle \rangle$) denotes nearest-neighbor 
(next-nearest-neighbor) bonds on the triangular lattice, defined by the basis vectors ${\bf a}_1=(1,0)$ and ${\bf a}_2=(1/2,\sqrt{3}/2)$. Periodic-boundary conditions 
on clusters with $N=L \times L$ ($L$ even) sites are considered, i.e., defined by ${\bf T}_j=L {\bf a}_j$, with $j=1,2$. We fix $J_1=1$ and $J_2=1/8$, where there 
is consensus that the ground state is a quantum spin liquid~\cite{kaneko2014,iqbal2016,hu2019}.

In the parton approach, the spin operator is expressed in terms of fermionic creation and annihilation operators:
\begin{equation}\label{eq:parton}
S^\alpha_i = \frac{1}{2} \sum_{\tau,\tau^\prime} c^\dag_{i,\tau} \sigma^\alpha_{\tau,\tau^\prime} c^\dagga_{i,\tau^\prime},
\end{equation}
where $(\sigma^x,\sigma^y,\sigma^z)$ are Pauli matrices. Within this framework, the Hilbert space is enlarged, from two to four states per site. Therefore, an 
additional constraint fixing one fermion per site must be imposed, in order to describe the original Hilbert space of $S=1/2$ degrees of freedom. At this stage,
the spin Hamiltonian~\eqref{eq:ham} is quartic in the fermionic operators and cannot be solved in general; still, in the spirit of a mean-field decoupling, a
quadratic (auxiliary) Hamiltonian can be defined, whose eigenstates determine the variational wave functions. In the simplest case, the auxiliary model includes 
only (spin-diagonal) hopping terms:
\begin{equation}\label{eq:aux_ham}
{\cal H}_{\text{aux}} = \sum_{\langle i j \rangle, \sigma} \chi_{ij} c^\dag_{i,\sigma} c^\dagga_{j,\sigma} + \text{h.c.},
\end{equation}
where $\chi_{ij}$ is the hopping amplitude that, in the general case, can be complex, implying the presence of a magnetic field in the auxiliary Hamiltonian. 
The magnetic flux through a given plaquette is determined by the argument of the product of hoppings around it. Notice that the unitary (gauge) transformation 
$c^\dagga_{j,\sigma} \to e^{i\theta_j} c^\dagga_{j,\sigma}$ on the site $j$ modifies the phase of the hopping amplitude of the six bonds emanating from $j$, 
without changing the magnetic flux through each of the $2N$ elementary (triangular) plaquettes, and the paths around the cluster along ${\bf a}_1$ and ${\bf a}_2$.

In general, eigenstates are easily obtained by a matrix diagonalization; then, a many-body state $\ket{\Phi}$ with $N$ fermions is attained by filling suitable 
single-particle states, e.g., the lowest-energy ones. Since the hopping amplitudes $\chi_{ij}$ do not depend on the spin, the spatial part of the eigenvectors 
can be chosen to be the same for $\sigma=\uparrow$ and $\downarrow$. Then, whenever both levels are occupied, the resulting many-body state is a spin singlet.

The constraint of one fermion per site is fulfilled by including the Gutzwiller projector:
\begin{equation}\label{eq:gutz}
{\cal P}_G=\prod_i (n_{i,\uparrow} - n_{i,\downarrow})^2,
\end{equation}
where $n_{i,\sigma}=c^\dag_{i,\sigma} c^\dagga_{i,\sigma}$ is the number operator on site $i$ (for spin $\sigma$). As a result, a suitable wave function for the 
original spin Hamiltonian is formally given by:
\begin{equation}\label{eq:spinwf}
\ket{\Psi}= {\cal P}_G \ket{\Phi},
\end{equation}
whose variational energy is straightforwardly evaluated within a quantum Monte Carlo scheme~\cite{beccabook}.

\begin{figure}
\includegraphics[width=0.9\columnwidth]{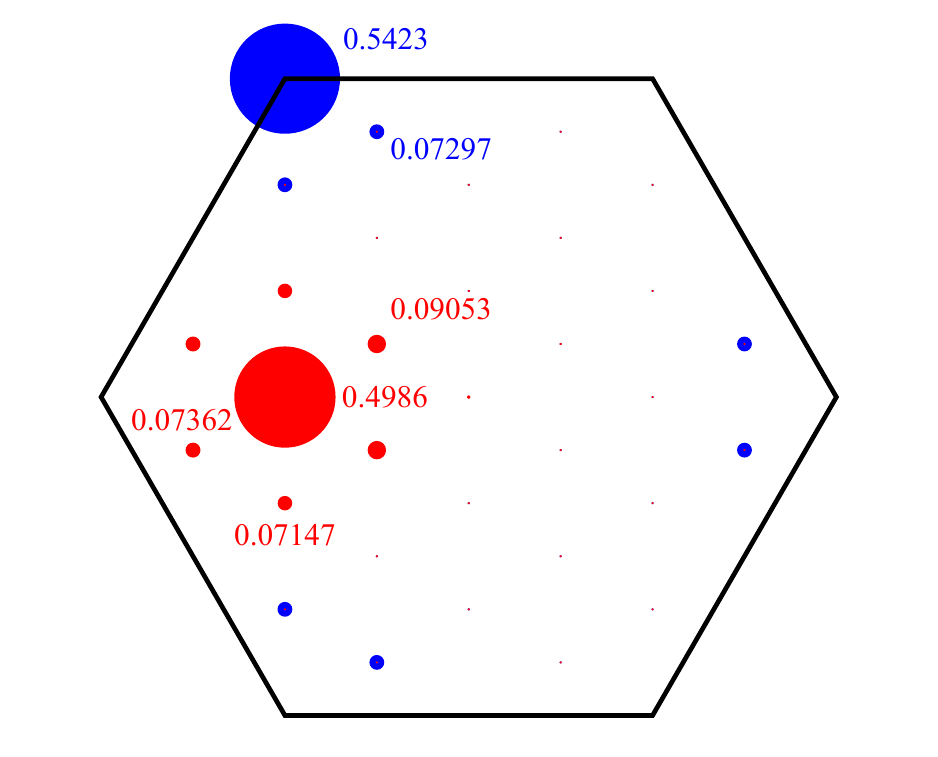}
\caption{\label{fig:ED}
Projection of the Gutzwiller-projected monopoles onto different subspaces with fixed momentum for the $6\times6$ cluster. The radius of the circles are proportional 
to the modulus (squared) of the projection and the largest values are explicitly reported. The blue circles [centered around ${\bf k}=(-2\pi/3,2\pi/\sqrt{3})$]
refer to the triplet monopole, while the red circles [centered around ${\bf k}=(-2\pi/3,0)$] pertain to one out of the three singlet monopoles. All the other 
states give similar results.}
\end{figure}

\section{Symmetries of the wave functions}\label{sec:symmetry}

\subsection{The ground state}
 
An accurate variational approximation of the ground state is obtained by filling the lowest-energy states of the auxiliary Hamiltonian~\eqref{eq:aux_ham} with
$\chi_{ij}=\pm 1$, so to have $\pi$ ($0$) magnetic fluxes on downward (upward) triangular plaquettes~\cite{lu2016,iqbal2016}, see Fig.~\ref{fig:LxLdirac}. 
In addition, periodic or antiperiodic boundary conditions along each direction can be imposed, thus giving four possible choices of the boundary conditions. 
The flux pattern of Fig.~\ref{fig:LxLdirac} is realized by breaking the translational symmetry in the auxiliary Hamiltonian (as in the Hofstadter 
problem~\cite{hofstadter1976}), giving rise to a periodic lattice with a $2 \times 1$ unit cell in the simplest Landau gauge. Notably, in the thermodynamic 
limit, the fermionic (unprojected) spectrum possesses two Dirac cones per spin. Therefore, the Gutzwiller-projected wave function is dubbed as $U(1)$ Dirac 
spin liquid. 

Whenever the unprojected wave function is a singlet, reversing the fluxes through upward and downward triangles does not change the Gutzwiller-projected state. 
In fact, the particle-hole transformation $P_h$ (i.e., $c^\dagga_{j,\sigma} \to c^\dag_{j,\sigma}$) changes sign to the auxiliary Hamiltonian, thus reversing the 
flux pattern; then, if $\ket{\Phi}$ is an eigenstate corresponding to one choice of fluxes, then $P_h \ket{\Phi}$ is an eigenstate for the reversed one (and 
identical boundary condtions). Moreover, ${\cal P}_G P_h \ket{\Phi}$ coincides with ${\cal P}_G \ket{\Phi}$, because ${\cal P}_G$ commutes with $P_h$ and the 
effect of the latter one is simply to reverse all the spins; but, in a singlet state, the spin flip leaves the state unaltered.

Let us now discuss the conditions under which the previous wave function is unique. The inversion symmetry $I$ swaps up and down triangles, leaving the boundary 
conditions unchanged. Then, the composite transformation $\tilde{I}={\cal G}_{I} I$, defined as the inversion followed by a suitable gauge transformation 
${\cal G}_{I}$, changes ${\cal H}_{\text{aux}} \to -{\cal H}_{\text{aux}}$, i.e., it anticommutes with the auxiliary Hamiltonian. As a consequence, for every 
single-particle level $\epsilon$, an energy level $-\epsilon$ is also present and $\tilde{I}$ brings one eigenstate $\ket{\phi}$ into the other. Furthermore, 
$\tilde{I}^2=1$, implying that $\tilde{I}$ has eigenvalues $\pm 1$. For $\epsilon \ne 0$, it is possible to define two states $\ket{\phi} \pm \tilde{I} \ket{\phi}$ 
with opposite eigenvalues of $\tilde{I}$; therefore, a different number of $+1$ and $-1$ eigenvalues implies the existence of zero-energy particle states. Then,
we focus on the inversion operator $\tilde{I}$, whose eigenstates can be constructed from states in which electrons are localized on sites of the lattice (and 
have a definite spin value), $|{\bf R},\sigma\rangle$. Let us consider a $L \times L$ cluster, where the hopping pattern is shown in Fig.~\ref{fig:LxLdirac} and
the coordinates of the sites are denoted by ${\bf R}=n_1 {\bf a}_1+n_2 {\bf a}_2$. Here, the gauge transformation ${\cal G}_I$ is simply given by the phase factor
$e^{i\pi(n_1+n_2)}$ at each site. For this geometry, there are exactly four sites that are left unchanged by the inversion, i.e., the ones with $(n_1,n_2)=(0,0)$,
$(L/2,0)$, $(0,L/2)$, and $(L/2,L/2)$. A pair of eigenstates of $\tilde{I}$ with opposite eigenvalues are given by $|{\bf R},\sigma\rangle \pm |-{\bf R},\sigma\rangle$ 
for each of the $(N-4)/2$ pairs of lattice sites ${\bf R}$ and $-{\bf R}$, which are swapped by the inversion operator. Instead, the states $|{\bf R},\sigma\rangle$,
which are left unchanged under inversion, are already eigenstates of $\tilde{I}$ with eigenvalue determined by the gauge transformation ${\cal G}_I$ and by the phase 
coming from the chosen boundary condition in the auxilary Hamiltonian. The eigenvalues of these four cases are reported in Table~\ref{tab:signs}. For periodic-periodic 
boundary conditions with $L=4n$ or antiperiodic-antiperiodic boundary conditions for $L=4n+2$, exactly $N+4$ eigenstates of $\tilde{I}$ with eigenvalue $+1$ are 
present. This implies that, in these two cases, eight zero-energy states (taking into account the spin degeneracy) necessarily belong to the spectrum, preventing 
the possibility to have a unique ground state of the auxiliary Hamiltonian at half filling. For all the other choices of boundary conditions, the argument shows that 
there are exactly $N$ eigenstates of $\tilde{I}$ with eigenvalue $+1$, suggesting that the are no zero energy states in the single-particle spectrum. This fact can 
be verified by performing the diagonalization of the auxiliary Hamiltonian. 

\begin{table}
\begin{tabular}{c|c|c|c|c}
\hline
$(n_1,n_2)$   & PBC-PBC         & PBC-APBC        & APBC-PBC        & APBC-APBC       \\
\hline
$(0,0)$       & $+1$            & $+1$            & $+1$            & $+1$            \\
$(L/2,0)$     & $e^{iL \pi/2}$  & $e^{iL \pi/2}$  & $-e^{iL \pi/2}$ & $-e^{iL \pi/2}$ \\
$(0,L/2)$     & $e^{iL \pi/2}$  & $-e^{iL \pi/2}$ & $e^{iL \pi/2}$  & $-e^{iL \pi/2}$ \\
$(L/2,L/2)$   & $+1$            & $-1$            & $-1$            & $+1$            \\
\hline
\end{tabular}
\caption{\label{tab:signs} The eigenvalues of the $\tilde{I}$ operator of the four states that are left unchanged by the inversion for the different choices
of the boundary conditions. The $L \times L$ cluster (with even $L$) is shown in Fig.~\ref{fig:LxLdirac} and the coordinates of the sites  are denoted by 
${\bf R}=n_1 {\bf a}_1+n_2 {\bf a}_2$.}
\end{table}

Although the fermionic Hamiltonian breaks the translational symmetry of the lattice, translational invariance is restored after Gutzwiller projection, because 
both elementary translations along the primitive vectors ${\bf a}_1$ and ${\bf a}_2$ leave all the fluxes unchanged and do not modify the boundary conditions. 
This means that, after a translation, it is possible to recover the original hopping pattern via a suitable gauge transformation. Therefore, we define the 
magnetic translation operators $\tilde{T}_j = {\cal G}_{T_j} T_j$. These symmetry operators commute with ${\cal H}_{\text{aux}}$ (although they do not commute with 
each other) implying that, whenever the many particle ground state of ${\cal H}_{\text{aux}}$ is unique, it must also be an eigenstate of both $\tilde{T}_j$. 
Since the gauge transformation commutes with the Gutzwiller projection and can be always chosen to act as the identity on $\ket{\Psi}$, the Gutzwiller-projected 
states are then zero-momentum eigenstates of the translation operators $T_j$. This is actually the case for three out of four possible boundary conditions for 
which no zero-energy modes are present, showing that we can find three linearly independent variational states.  

Finally, we consider the elementary rotation $R$ by $\frac{2\pi}{6}$ around the origin. This symmetry operation changes by $\pi$ all the fluxes through the 
elementary triangles because it swaps the up and down triangles. Moreover it also performs a cyclic permutation of the three boundary conditions. At half filling, 
if we combine the rotation with the previously defined particle-hole transformation $P_h$, the Hamiltonian acquires a further global minus sign and then the 
fluxes through the elementary triangles go back to their original value without affecting the boundary conditions. This means that, by performing a suitable 
gauge transformation ${\cal G}_{R}$, the combined symmetry $P_h R {\cal G}_{R}$ acts as a permutator on the three states. Therefore, in order to build a variational 
state that is eigenstate of the rotation operator, an appropriate linear combination of these three states must be constructed.

\begin{figure*}
\includegraphics[width=0.49\textwidth]{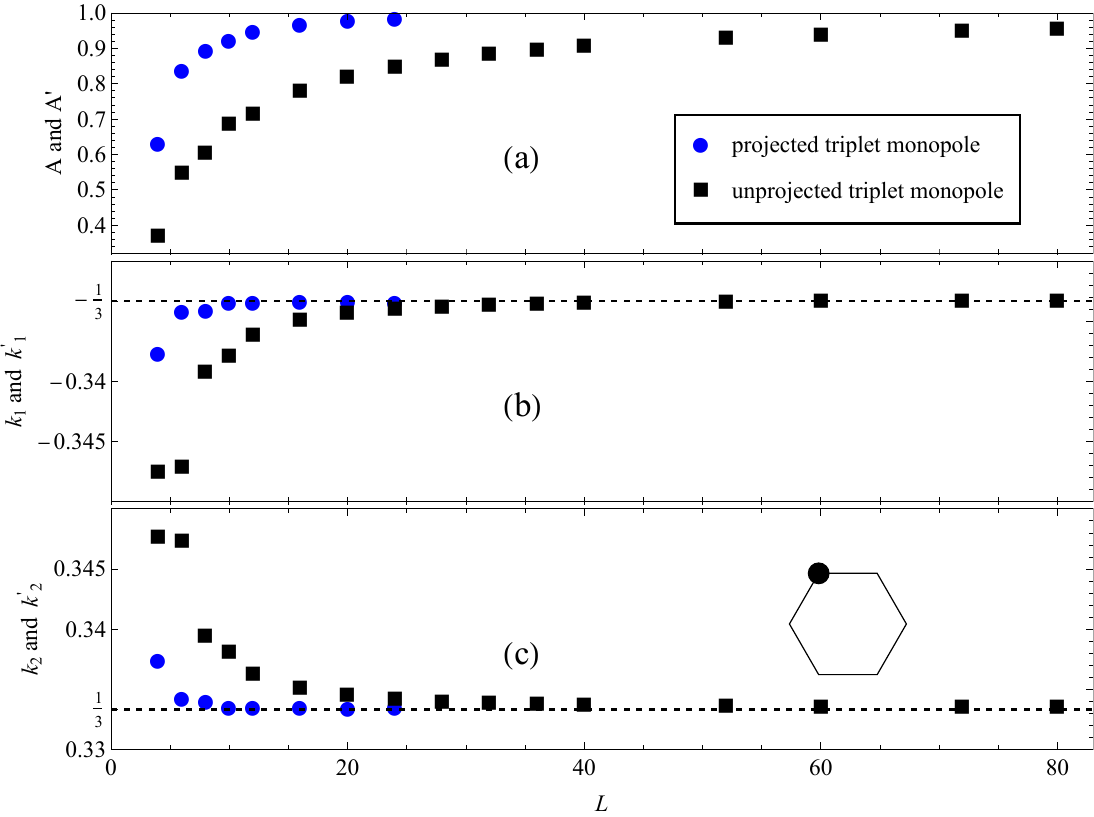}
\hfill
\includegraphics[width=0.49\textwidth]{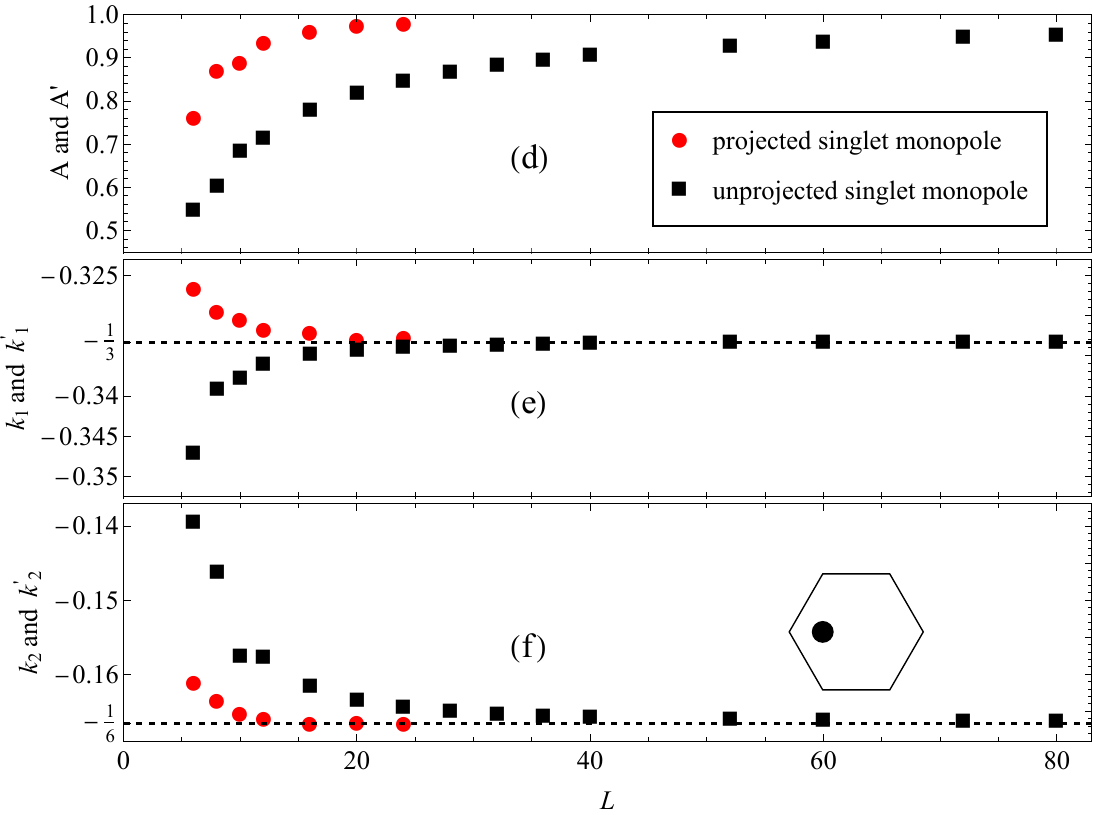}
\caption{\label{fig:momenta}
Amplitude and phase of unprojected and Gutzwiller-projected states (see Eqs.~\eqref{eq:mf_momentum} and~\eqref{eq:pro_momentum}, respectively) for the triplet (left
panels) and one of the three singlet monopoles (right panels) as a function of linear system size $L$. The insets show the points $k_j={\bf a}_j \cdot {\bf k}$ 
(in unit of $2\pi$) on the Brillouin zone corresponding to the plots. The error-bars of the Monte Carlo sampling are smaller than the size of the symbols.}
\end{figure*}

\subsection{The $Q=1$ monopole excitations}

The $Q=1$ monopole excitations are constructed by inserting a $2\pi$ flux on top of the Dirac state, spreading it uniformly through all the elementary plaquettes 
of the lattice~\cite{song2019,song2020,budaraju2023}. Inserting a $-2\pi$ flux would yield an antimonopole, whose wave function is the complex conjugate of the 
monopole one. One possible choice for the fermionic hoppings is shown in Fig.~\ref{fig:LxLmonopole}, where the additional flux is equally distributed in the two 
triangles within each unit cell. It should be noted that, on a closed manifold such as the torus, the existence of a monopole with total flux $2\pi$ is always
accompanied by a $-2\pi$ flux, which is effectively hidden (modulo $2\pi$) in a single plaquette (e.g., in our setup, at the top-right corner). Most importantly, 
the pattern of the hopping amplitudes is not translationally invariant and it is not possible to restore the symmetry by using a gauge transformation (unlike for 
the Dirac state discussed above). Indeed, the (gauge invariant) fluxes along the loops wrapping around the torus are not uniform, which are not left invariant 
upon translation (in addition, different choices for the boundary conditions do not lead to different states). The Gutzwiller-projected states are also not 
translationally invariant. Still, it is possible to find a gauge transformation ${\cal G}^{m}_{T_j}$ for which the hopping amplitudes that determine the symmetry 
breaking are confined along lines or columns of $O(L)$ bonds~\cite{song2019}. As a result, translational symmetries are restored in the thermodynamic limit.

On any finite cluster, the average momentum of the unprojected wave function is evaluated by:
\begin{equation}\label{eq:mf_momentum}
\frac{\bra{\Phi} {\cal G}^{m}_{T_j} T_j \ket{\Phi}}{\braket{\Phi|\Phi}} = A e^{ik_j},
\end{equation}
which can be easily computed analytically. Instead, the average momentum of the Gutzwiller-projected state is defined by:
\begin{equation}\label{eq:pro_momentum}
\frac{\bra{\Psi} T_j \ket{\Psi}}{\braket{\Psi|\Psi}} = A^\prime e^{ik^\prime_j},
\end{equation}
which may be evaluated by using the variational Monte Carlo approach. A wave function with a well defined momentum ${\bf k}$ (with $k_j={\bf a}_j \cdot {\bf k}$) 
is signalled by an amplitude $A=1$ in the unprojected case, or $A^\prime=1$ in the projected one.

The monopole insertion modifies the single particle spectrum of the auxiliary Hamiltonian~\eqref{eq:aux_ham}, whose properties can be again determined by examining 
the effect of inversion symmetry $I$. Both the additional flux coming from the monopole and the boundary conditions are unchanged under inversion and, therefore, 
it is again possible to find a suitable gauge transformation ${\cal G}^{m}_{I}$ such that $\tilde{I}={\cal G}^{m}_{I} I$ just changes sign to the auxiliary 
Hamiltonian. This proves that, also in this case, the single particle spectrum is symmetrical and that $\tilde{I}$ takes an eigenstate of energy $\epsilon$ into 
an eigenstate of energy $-\epsilon$. The only difference induced by the presence of the monopole is that now the operator $\tilde{I}$ has just $N-2$ eigenvalues 
equal to $-1$ and $N+2$ eigenvalues equal to $+1$ (or {\it viceversa}), implying that four (two per spin) single-particle energy levels must have a vanishing energy. 
These four levels must be occupied by two fermions, in order to have $N$ fermions in the many-body state. Therefore, three singlets and one triplet can be obtained.
In particular, filling the zero-energy levels by two fermions with the same spin (either up or down) gives two closed-shell configurations with $S^z=\pm 1$. Since, 
in both cases the many-body state is unique, the momentum turns out to be well defined in the thermodynamic limit. By contrast, filling these levels with fermions 
having opposite spins leads to four (open-shell) states with $S^z=0$, i.e., $\ket \Phi_{\alpha}$ with $\alpha=1,\dots 4$ (the corresponding Gutzwiller-projected 
wave functions are denoted by $\ket \Psi_{\alpha}$). In order to construct the approximate eigenstates of translations on finite clusters, we diagonalize a pair 
of $4 \times 4$ matrices with elements $\bra{\Phi_{\alpha}} {\cal G}^{m}_{T_j} T_j\ket{\Phi_{\beta}}$ (with $j=1,2$) for the unprojected case (and 
$\bra{\Psi_{\alpha}} T_j\ket{\Psi_{\beta}}$ for the projected one). The eigenstates of these matrices separate into two blocks of dimensions $3 \times 3$ and 
$1 \times 1$ (corresponding to the singlet and triplet sectors, respectively), thus allowing us to extract the information about the momenta. The triplet has 
momentum ${\bf k}=(-2\pi/3,2\pi/\sqrt{3})$ and the singlets have ${\bf k}=(-2\pi/3,0)$ and ${\bf k}=(\pi/3,\pm \pi/\sqrt{3})$. Indeed, the total wave function 
can be factorized, splitting fermions with up and down spin. Within the adopted choice of the hoppings (see Fig.~\ref{fig:LxLmonopole}), the many-body state that 
is obtained by filling all the strictly negative energy orbitals (i.e., $N/2-1$ levels for each spin value) has momentum $(-2\pi/3,2\pi/\sqrt{3})$ (the two spin 
components contribute equally). Then, two additional fermions (with opposite spins, which is suitable for both triplet and singlet states) must be added on the 
zero-energy states. This can be done in four possible ways. We already showed that the zero energy eigenstates are also eigenstates of the modified inversion 
operator ${\tilde I}$ and, therefore, the momenta of any two particle eigenstate of ${\tilde T}_i$ must be invariant upon changing ${\bf k} \to -{\bf k}$. This 
implies that, for these states, $\langle {\tilde T}_i\rangle = \pm 1$. The results discussed previously correspond to the triplet state (in the thermodynamic 
limit), with $\langle {\tilde T}_1\rangle = \langle {\tilde T}_2\rangle=1$, and the three singlets, with $\langle {\tilde T}_1\rangle = -\langle {\tilde T}_2\rangle$
and $\langle {\tilde T}_1\rangle = \langle {\tilde T}_2\rangle=-1$. This analysis is consistent with the calculations reported in Ref.~\cite{song2019}. A similar 
construction applies also for the antimonopole, leading to an additional triplet and three singlets, with opposite momenta ${\bf k} \to -{\bf k}$; as a result, 
a total of two triplet and six singlet states are constructed. 

\section{Results}\label{sec:results}

\subsection{Exact diagonalizations}

Here, we briefly report exact calculations on the $6 \times 6$ cluster. Let us start with the ground-state calculations. As discussed in Sec.~\ref{sec:symmetry}, 
the three choices of boundary conditions (i.e., excluding the antiperiodic-antiperiodic one) give a unique ground-state wave function with $N$ fermions occupying 
the lowest-energy single-particle states of the auxiliary Hamiltonian~\eqref{eq:aux_ham} (which has a finite-size gap in the spectrum). From these many-body 
states, it is possible to construct three wave functions with all the symmetries of the original (spin) Hamiltonian, i.e., with zero momentum and eigenvalues 
of the rotation operator $R$ equal to $1$ and $e^{\pm i\frac{2\pi}{3}}$. The former one corresponds to the lowest-energy variational {\it Ansatz}, with energy 
$E_{0}/J_1=-18.271$ and overlap with the exact ground state equal to $0.929$. The latter two (degenerate) states have energy $E/J_1=-17.688$ and overlap with 
the second lowest-energy state in the same subspace equal to $0.803$. We mention that our construction for the ground-state wave function (namely, the linear 
superposition of three different boundary conditions) is different from what has been taken in Ref.~\cite{wietek2024}, where a {\it single} complex boundary
condition is considered. Still, both the variational energy and its overlap with the exact ground state are comparable.

We now show the results for the $Q=1$ monopoles. Since, the monopole state has not a well defined momentum on the $6 \times 6$ lattice, we define a (non-normalized)
state by the projection ${\cal P}_k$ on a given subspace with fixed ${\bf k}$. The resulting squared norm is given by:
\begin{equation}
{\cal N}^2_k = \bra{\Psi} {\cal P}^2_k \ket{\Psi},
\end{equation}
where $\ket{\Psi}$ is either a triplet or a singlet monopole state (for the singlets, the diagonalization of the $4 \times 4$ matrix is employed to define suitable
{\it Ans\"atze}). We have that $\sum_k {\cal N}^2_k=1$. The results are shown in Fig.~\ref{fig:ED}: in all cases, the norm is relatively large for the expected 
momenta, i.e., ${\bf k}=(-2\pi/3,2\pi/\sqrt{3})$ for the triplet and ${\bf k}=(-2\pi/3,0)$ (or equivalently ${\bf k}=(\pi/3,\pm \pi/\sqrt{3})$) for the singlet. 
Moreover, the monopole wave functions have a large overlap with the exact lowest-energy eigenstates of the $J_1-J_2$ Heisenberg model, e.g., $0.69$ and $0.62$ 
for the triplet and singlet cases, respectively. These results are in agreement with those obtained in Ref.~\cite{wietek2024}. We mention that the normalized 
monopole states (i.e., $\ket{\tilde{\Psi}_k}={\cal P}_k\ket{\Psi}/{\cal N}_k$) the overlaps become $0.91$ and $0.87$, better emphasizing the accuracy of the parton 
approach in constructing these excitations.

\begin{figure}
\includegraphics[width=\columnwidth]{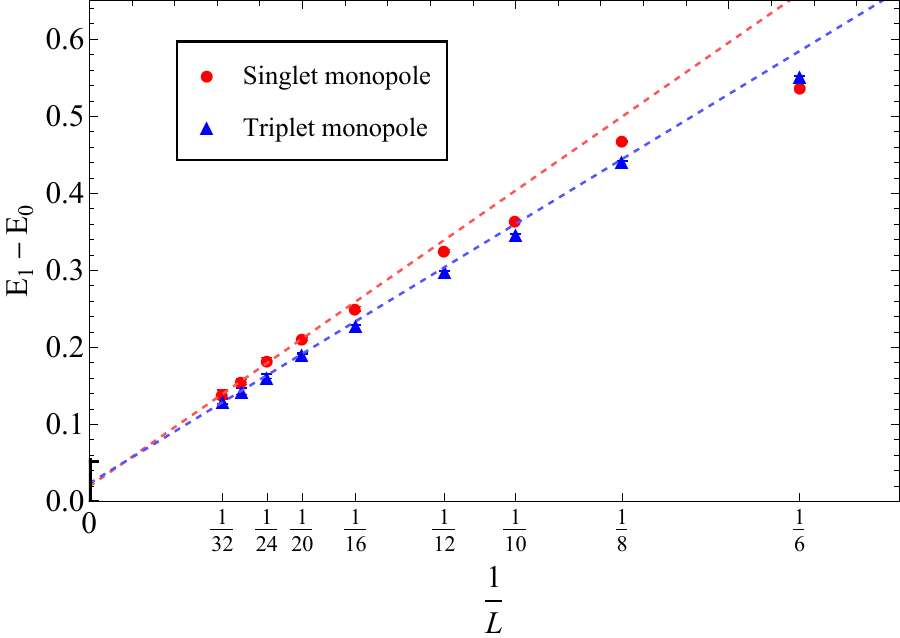}
\caption{\label{fig:scaling}
Monopole gap $E_{1}-E_{0}$ on the triangular lattice as a function of $1/L$. Both triplet and singlet cases are reported.}
\end{figure}

\begin{figure}
\includegraphics[width=\columnwidth]{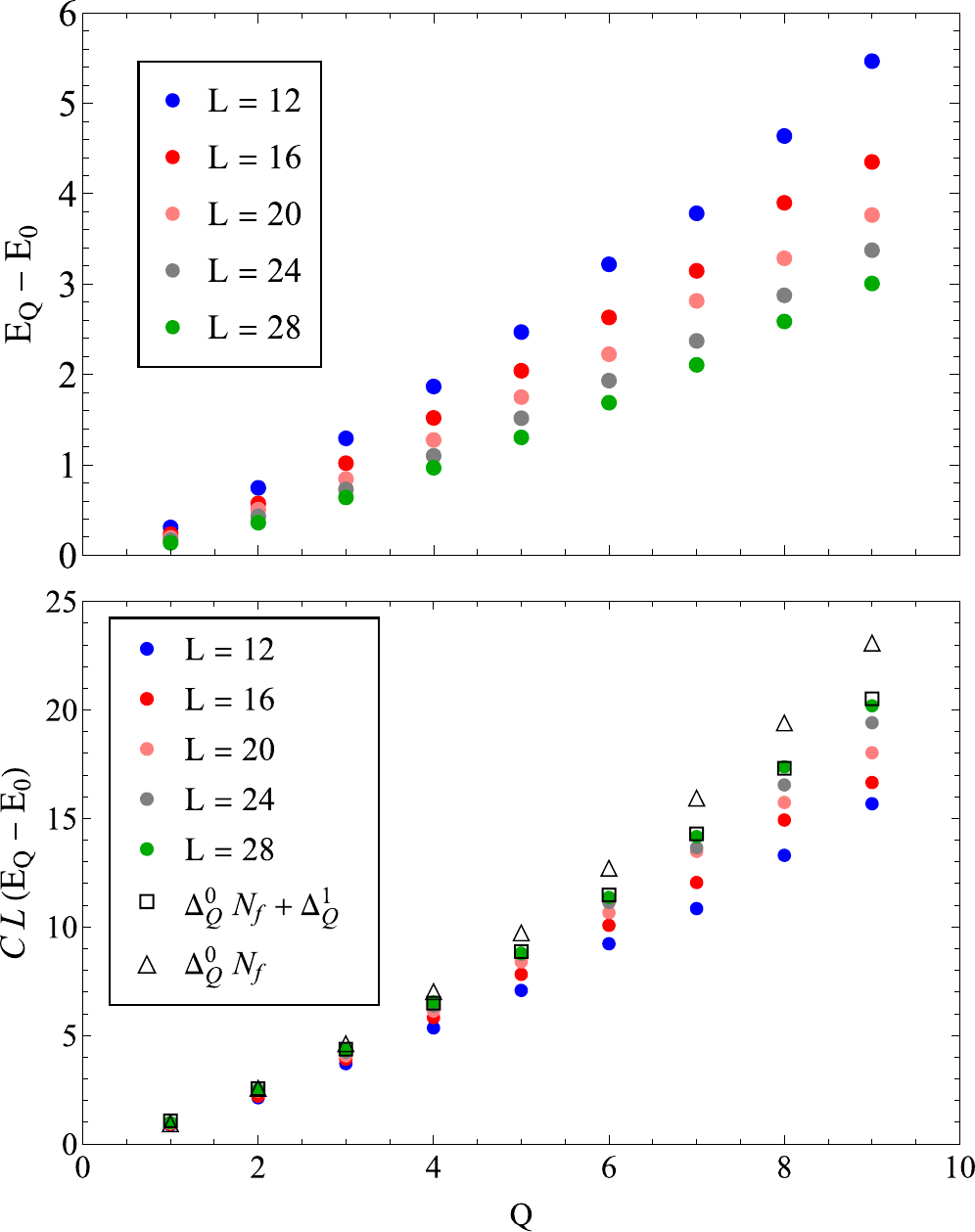}
\caption{\label{fig:manymono}
Upper panel: Variational energy of charge-$Q$ (singlet) monopole excitations $E_{Q}-E_{0}$, as a function of $Q$ for different sizes of the cluster $L$. Lower panel: 
the excitations energies multiplied by $C \times L$ (where $C=0.24$, see text) and compared with the scaling dimensions obtained by field-theoretical calculations in 
the limit of a large number of Dirac points in the fermionic spectrum. In particular, we report both the leading-order contribution proportional to $N_f$ (set to 4) 
and the $O(1)$ correction~\cite{borokhov2002,dyer2013,dupuis2022}.}
\end{figure}

Interestingly, the monopole excitations have a large overlap with particle-hole spinon excitations, obtained by starting from the Dirac state (with the hopping 
amplitudes given in Fig.~\ref{fig:LxLdirac}) and allowing a different occupation of the fermionic levels. This approach, has been recently considered to assess 
the dynamical structure factor~\cite{ferrari2019}. For the $6 \times 6$ cluster, we can improve this construction, by including states with three different 
boundary conditions (in Ref.~\cite{ferrari2019}, we only considered one of them). For the triplet case, starting from the Dirac unprojected state, we build all 
the possible particle-hole spinon excitations at ${\bf k}=(-2\pi/3,2\pi/\sqrt{3})$ after projection. Then, in the $27$-dimensional subspace (corresponding to
$9$ linearly independent states for each boundary condition) spanned by these spinon states, we find the optimal linear combination minimizing the energy (after 
projection). The spinon variational energy at ${\bf k}=(-2\pi/3,0)$ is $E_{\rm sp}/J_1=-17.762$, while the monopole variational energy is $E_{1}/J_1=-17.817$. 
Remarkably, the overlap between the monopole and the spinon states is $0.67$ (or $0.92$ for the normalized monopole state). Similarly, for the singlet monopole, 
the spinon variational energy is $E_{\rm sp}/J_1=-17.776$, to be compared with the monopole one $E_{1}/J_1=-17.694$. The overlap between these two states is $0.64$ 
(or $0.90$). These results suggest that bare spinons and monopoles do not exist as free particles and cannot be disentangled in the exact eigenfunctions of the 
frustrated Heisenberg model.

\subsection{Variational Monte Carlo}

Let us now consider larger clusters, for which the variational Monte Carlo approach is needed, and focus on the average momentum of Eq.~\eqref{eq:pro_momentum}.
In Fig.~\ref{fig:momenta}, we show the behavior of the absolute value ($A^\prime$) and phases ($k^\prime_i$) as a function of system size for both the triplet 
and one singlet, in comparison to the unprojected values ($A$ and $k_i$), evaluated by Eq.~\eqref{eq:mf_momentum}. As expected, all the absolute values converge 
to $1$ in the thermodynamic limit. However, we observe that the convergence in presence of the Gutzwiller projection is much faster than in the unprojected cases. 
For example, $A^\prime$ for the projected triplet monopole at $L=16$ already exceeds $A$ for the unprojected one at $L=80$. Obviously, since the Gutzwiller
projector commutes with the translations, the thermodynamic values of the momenta of the projected states coincide with the ones obtained within the unprojected 
calculations. The above results are consistent with the symmetry analysis of the inversion operator discussed in the previous section. 

\begin{figure*}
\includegraphics[width=\textwidth]{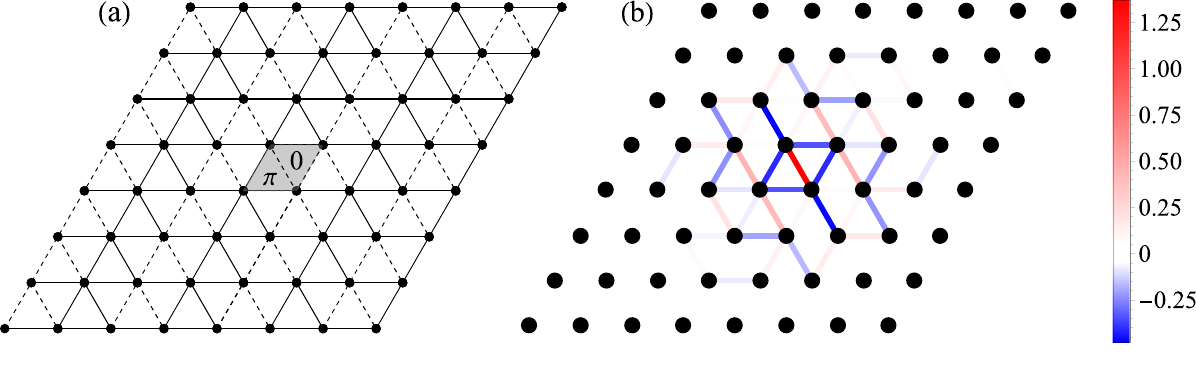}
\caption{\label{fig:loc_mono_setup}
(a) The hopping amplitudes $\chi_{ij}$ of the Dirac {\it Ansatz} with the insertion of two localized gauge fluctuations in the rhomboid (shaded) plaquettes on 
the cluster with $L=8$. Solid and dashed lines denote the hoppings $\chi_{ij}=1$ and $-1$, respectively. (b) Rescaled nearest-neighbor energies ${\cal S}(i,j)$ 
of Eq.~\eqref{eq:spinspin} for the Gutzwiller-projected wave function on the same cluster.}
\end{figure*}

Let us now move to the energetics of monopole excitations. The results for triplet and singlet monopole gaps $E_{1}-E_{0}$ (where $E_{0}$ and $E_{1}$ are the 
variational energies of the Dirac state and the charge $Q=1$ monopole one, respectively) are shown in Fig.~\ref{fig:scaling} as a function of inverse system size 
$1/L$. In order to avoid complicated Monte Carlo calculations with multi-determinant states, the singlet state is constructed by occupying with two fermions (with 
opposite spins) the same zero-energy level (not by the diagonalization of the $4 \times 4$ matrix, as discussed before); we verified that, on small clusters, this 
simple approach yields the same energy (within error-bars) as the calculation involving a linear superposition of different occupations of the zero-energy levels. 
In the thermodynamic limit, monopole excitations become gapless, supporting the existence of a $U(1)$ Dirac spin liquid in the frustrated region of the $J_1-J_2$ 
Heisenberg model on the triangular lattice. 

In addition, we consider higher values of the monopole charge $Q$, i.e., including a magnetic flux of $2\pi Q$. On finite clusters, for $Q \ll N$, there are 
$2Q$ zero-energy levels per spin (in some cases, the energies are not exacly zero, but still separated by the other ones). These states must be filled with $2Q$ 
fermions (with up or down spin), thus increasing the complexity for constructing many-body wave functions with a definite spin value. Then, we take a simplified 
approach and construct states with fermions occupying (with opposite spins) the same zero-energy level, thus leading to a singlet. We remark that no differences 
in the energy of Gutzwiller-projected states are detected by occupying different single-particle levels. The energy of charge-$Q$ monopoles is expected to scale 
with a non-trivial behavior, as reported in the field-theoretical calculations of Ref.~\cite{borokhov2002} for the leading-order term in the $1/N_{f}$ expansion. 
Our results for $E_{Q}-E_{0}$ are reported Fig.~\ref{fig:manymono} for a few values of $Q$ and different cluster sizes $L$. In addition, we also report the direct 
comparison between our variational results multiplied by $L$ and the scaling dimensions obtained by field-theoretical calculations performed within the 
large-$N_{f}$ expansion. Setting $N_f=4$, we consider both the leading contribution ($\Delta^0_Q N_f$)~\cite{borokhov2002} and the first-order correction 
($\Delta^0_Q N_f + \Delta^1_Q$)~\cite{dyer2013,dupuis2022}. However, this comparison requires a normalization factor due to the different geometries (the torus 
and the conformal sphere, in the variational and analytical approaches, respectively) and the different units adopted in the two cases (e.g., $J_1=1$ and 
$\hbar=c=1$, in the variational and analytical approaches, respectively). Due to these uncertainties in the evaluation of the ``correct'' normalization factor, 
we chose to multiply our results by $C=0.24$ (which is obtained by matching the exact results with first-order correction with our $L=28$ calculations).
This outcome gives further support for the stability of the Dirac spin liquid in the $J_1-J_2$ Heisenberg model on the triangular lattice and the fact that 
the corresponding quantum spin liquid is described by quantum electrodynamics in $2+1$ dimensions. We further notice that in a previous work on the Kagome 
lattice~\cite{budaraju2023}, we argued that the energy density $(E_Q - E_0)/L^2$ is a function of the monopole density $Q/L^2$ in the thermodynamic limit: this 
is consistent with the scaling form $(E_Q - E_0) \approx Q^{3/2}/L$ for large $Q$ values, as given by Ref.~\cite{borokhov2002}. In fact, we have checked that also 
the results of our previous work~\cite{budaraju2023} can be plotted in the same way as Fig.~\ref{fig:manymono} with a very good collapse to field theory predictions.
This universal nature of the energy spectrum on the torus was also discussed in Ref.~\cite{schuler2016} for the Ising conformal-field theory.

Within the monopole construction, the total gauge flux $2\pi Q$ is spread uniformly in each plaquette, thus affecting the structure of the fermionic hopping in 
the whole lattice. Still, local gauge fluctuations are expected to exist at any lengthscale, even though they might not be part of the low-energy spectrum. The 
simplest case that can be considered is given by concentrating the full ($2\pi$) magnetic flux in a single rhomboidal plaquette, similarly to vison excitations 
in $\mathbb{Z}_2$ spin liquids~\cite{kitaev2006}. Indeed, whenever a proximate $\mathbb{Z}_2$ state exists, a vestige of such states should be present at relatively 
small (but finite) energies in the $U(1)$ spin-liquid phase.

The prototypical example of such localized gauge excitations is shown in Fig.~\ref{fig:loc_mono_setup}. This configuration would correspond (in a $\mathbb{Z}_2$ 
spin liquid) to the creation of a couple of nearest-neighbor visons. The actual location of the defect in the pattern of fluxes may be detected by considering 
spin-spin correlations at nearest neighbors:
\begin{equation}\label{eq:spinspin}
{\cal S}(i,j) = \frac{\bra{\Psi_{\rm loc}} {\bf S}_i \cdot {\bf S}_j \ket{\Psi_{\rm loc}} - \bra{\Psi_{0}} {\bf S}_i \cdot {\bf S}_j \ket{\Psi_{0}}}
{\left| \bra{\Psi_{0}} {\bf S}_i \cdot {\bf S}_j \ket{\Psi_{0}} \right|},
\end{equation}
where $\ket{\Psi_{\rm loc}}$ is the state containing two localized fluxes and $\ket{\Psi_{0}}$ is the Dirac {\it Ansatz}. Notably the spin-spin correlations of 
$\ket{\Psi_{\rm loc}}$ differ from the ones of the Dirac state $\ket{\Psi_{0}}$ only in the vicinity of the perturbation, see Fig.~\ref{fig:loc_mono_setup}.

We observe that this variational state has a large overlap with the Dirac {\it Ansatz}, namely, $\ket{\Psi_{\rm loc}} = \alpha^2 \ket{\Psi_0} + (1-\alpha^2) 
\ket{\Psi_{\rm exc}}$, with $\alpha^2 \approx 0.85$ (weakly depending on the size of the cluster, for large $L$). This fact is strikingly different compared
to $\mathbb{Z}_2$ spin liquids. Then, from the direct calculation of the the variational energy of $\ket{\Psi_{\rm loc}}$ and the value of $\alpha^2$, we can 
directly estimate $(E_{\rm exc}-E_{0})/J_1 \approx 0.87$ in the thermodynamic limit. The present result gives a strong confirmation that this kind of states
have relatively high energy and are not particularly relevant for the low-energy properties of the $U(1)$ spin liquid.

\section{Conclusions}\label{sec:conclusions}

The $U(1)$ Dirac spin liquid is an exotic quantum state, characterized by low-energy excitations involving spinons and monopoles. In this work, we thoroughly 
examined monopole excitations relevant to the low-energy physics of the $J_1-J_2$ Heisenberg model on the triangular lattice, employing the parton construction
supplemented by the Gutzwiller projection. Our key findings are: (i) in the thermodynamic limit, the charge $Q=1$ monopole wave functions acquire the predicted 
momenta~\cite{song2019} for both singlet and triplet cases (the convergence with $L$ is much faster for the Gutzwiller-projected states than for the unprojected 
ones); (ii) monopole excitations are gapless; (iii) on small clusters, both singlet and triplet monopoles have a large overlap with particle-hole spinon states,
suggesting that gauge fluctuations and spinons are strongly interacting; and (iv) there is a nice agreement between the variational energy of charge-$Q$ monopoles 
(on $L \times L$ tori) and the one obtained by field-theoretical approaches (on the conformal sphere)~\cite{borokhov2002,dyer2013,dupuis2022}. Finally, we 
investigated localized gauge excitations on individual plaquettes (similar to visons in $\mathbb{Z}_2$ spin liquids). These kind of excitations have a large 
overlap with the Dirac wave function (and the orthogonal component corresponds to high-energy states). All these outcomes give confirmation that the (fermionic) 
parton construction captures the correct description of the $U(1)$ Dirac spin liquid. Our results serve to highlight the intricate physics of Dirac spin liquids 
in frustrated magnetic systems. From the experimental side, the detection of monopole excitations is subtle and may require some indirect probe. For example, it 
has been argued that they could lead to a spin-Peierls distortions a broadening/softening of certain phonon modes~\cite{seifert2024, ferrari2024}. Instead, from 
the numerical side, an intriguing direction for future exploration is to capture the phase transition out of the Dirac spin liquid phase through monopole 
proliferation. 

\section{Acknowledgements}
We thank L. Balents, S. Bhattacharjee, S. Capponi, and A. L\"auchli for stimulating discussions. F.B. particularly thanks L. di Pietro for important
explanations about the field theory in $2+1$ dimensions. S.B thanks G. Nambiar for discussions about the gauge undoing procedure. S.B, D.P and Y.I. acknowledge 
financial support from the Indo-French Centre for the Promotion of Advanced Research - CEFIPRA Project No. 64T3-1. D.P. acknowledges support from the TNTOP 
ANR-18-CE30-0026-01 grant awarded by the French Research Council. This work was granted access to the HPC resources of CALMIP center under allocations 2023-P1231 
and 2024-P1231. The work of Y.~I. was performed, in part, at the Aspen Center for Physics, which is supported by National Science Foundation Grant No.~PHY-2210452. 
The participation of Y.~I. at the Aspen Center for Physics was supported by the Simons Foundation (1161654, Troyer). This research was supported in part by grant 
NSF PHY-2309135 to the Kavli Institute for Theoretical Physics (KITP) and the International Centre for Theoretical Sciences (ICTS), Bengaluru through participating 
in the program - Kagome off-scale (code: ICTS/KAGOFF2024/08). Y.~I.\ acknowledges support from the ICTP through the Associates Programme, from the Simons Foundation 
through Grant No.~284558FY19, and IIT Madras through the Institute of Eminence (IoE) program for establishing QuCenDiEM (Project No.~SP22231244CPETWOQCDHOC). 
S.~B. and Y.~I acknowledge the use of the computing resources at HPCE, IIT Madras.


%

\end{document}